\input{aipcheck}

\documentclass[
    ,final            
  ]
  {aipproc}

\layoutstyle{8x11double}

\begin{document}

\title{Analysis of the Prompt Optical Emission of the Naked-Eye GRB 080319B}

\classification{98.70.Rz; 95.55.Cs; 95.75.Wx.}
\keywords      {$\gamma$-ray burst, ground-based observations, time variability.}

\author{C. Bartolini}{
  address={Astronomy Department of Bologna University, Bologna, Italy}
}

\author{G. Greco}{
  address={Astronomy Department of Bologna University, Bologna, Italy}
}

\author{A. Guarnieri}{
  address={Astronomy Department of Bologna University, Bologna, Italy}
}

\author{A. Piccioni}{
  address={Astronomy Department of Bologna University, Bologna, Italy}
}

\author{G. Beskin}{
  address={Special Astrophysical Observatory of Russian Academy of Science, Nizhnij
Arkhyz, Russia}
}

\author{S. Bondar}{
  address={Institute for Precise Instrumentation, Nizhnij Arkhyz, Russia}
}

\author{S. Karpov}{
  address={Special Astrophysical Observatory of Russian Academy of Science, Nizhnij Arkhyz, Russia}
}

\author{E. Molinari}{
  address={INAF-TNG, Canary Islands, Spain}
}

\begin{abstract}
  We present the observed$/$intrinsic optical parameters and the variability
  analysis of the Naked-Eye Burst, GRB 080319B, observed by the TORTORA
  wide-field optical monitoring system. The event is extreme not only in
  observed properties but also intrinsically: it is the most luminous event
  ever recorded at optical wavelengths.  The temporal properties suggest
  short-lived periodic activities of the internal engine.  This is the fastest
  optically variable source detected at cosmological distances.
\end{abstract}

\maketitle


\section{Introduction to the Phenomenon: Prompt Optical Emission}
Over the past nine years the prompt optical emissions contemporaneous with the
$\gamma$-ray-active phase of a GRB have became subject of exciting debate in
the astronomical community; the brevity of these phenomena and their fleeting
nature makes them elusive and difficult to study.  Due to the relative short
duration of the prompt $\gamma$-ray emission (T $\sim$ 0.2-100 sec)
simultaneous follow-up observations at optical wavelengths suffer lack of rapid
and precise burst localization.  GRB 990123 was the first event for which
optical emission was detected during the burst phase \citep{Ake99}.

Nowadays, the fast and accurate localization of GRBs by the Swift mission \citep{Geh2004} and
its capability to alert fast-slewing robotic
telescopes
within few seconds after the burst has allowed to significantly increase the
numbers of the events that are optically observed during the bursting phase,
although the prompt optical emissions are usually not well sampled.

During the last years, we have investigated different search strategies in  ultra fast photometry fields to measure
rapid changes in light intensity  in a phenomenon occurring within an extremely short period of time
and  randomly distributed  over the sky
\citep{picc}, \citep{Besk99}. The  high-speed and wide-field FAVOR \citep{KarpFAV} and TORTORA cameras
represent the more recent developments  in our search program in which  intensified CCD imaging
are employed and tested.
In particular the ground-based TORTORA  observations synchronized with  the $\gamma$-ray telescope   on board
of the Swift  satellite  (BAT: Burst Alert Telescope) has permitted to trace  the optical burst time structure of the GRB 080319B with
unprecedented level of accuracy.

At the beginning of the May 2006 TORTORA was installed in the dome of the REM
telescope located at ESO-La Silla Observatory (Chile) and successfully achieved
its First Light \citep{Mol06}.

\section{Observation and Data Reduction}
\subsubsection{Observation}
On 19 March 2008 at 06:12:49 UT (hereafter $t_{0}$) the Swift's Burst Alert
Telescope triggered and located GRB 080319B (trigger = 306757;
\citep{Rac08_GCN}) with a $\sim$ 3' radius error box.  The bright burst was
simultaneously detected by the Konus-Wind (KW) satellite \citep{Gol08} yielding
a burst fluence of $S_{\gamma}$ = 6.23 $\pm$ 0.13 $\times$ $10^{-4}$ $erg$
$cm^{-2}$ [20 keV to 7 MeV].

Assuming $z$ = 0.937 \citep{Vree08} and the standard cosmology model ($H_{0}$ =
71 $Km$ $s^{-1}$ $Mpc^{-1}$, $\Omega_{M}$= 0.27 and $\Omega_{\Lambda}$ = 0.73)
the isotropic energy release is $E_{iso}$ = $1.30 \times 10^{54} $ $erg$.

The field of the GRB 080319B was imaged before the GRB event by three
independently ground-based optical sky monitoring.  No optical precursors were
detected in TORTORA, "Pi of the Sky"  and RAPTOR
 surveys with observations starting 26 minutes, 16 seconds and 30
minutes before the Swift$/$BAT trigger, respectively.
In RAPTOR \citep{Woz08} and "Pi of the Sky" \citep{Cwi08} the first image with detectable optical emission
 started at  $t_{0}$ = 1.87 $\pm$ 5 sec and at $t_{0}$ =  2.75 $\pm$ 5 sec   after the BAT
trigger, respectively,  when the optical counterpart became brighter than V $\approx$ $10^{m}$.
In the TORTORA  high temporal resolution dataset (0.13 sec exposure time), the
first frame in which we detected the optical flux started at  $t_{0}$ = 9.18 $\pm$ 0.065 sec
after the BAT trigger when the source became brighter than V $\approx$ $8^{m}$ \citep{Kar08}.  The
bright visual peaks which occurred during the  prompt $\gamma$-ray emission
approximately reached V $\approx$ $5.5^{m}$, this made it visible with naked eye in the
BOOTES constellation\footnote{http://grb.sonoma.edu/.} for $\sim$ 40 seconds,
assuming an observer in a dark location.

Since 05:46:22 UT REM telescope had observed the box of previous burst, GRB
080319A. At 06:12:49 UT, Naked-Eye Burst, GRB 080319B, flashed at $\sim$10
degrees from the former, near the edge of TORTORA's field of view. At 06:13:13
UT, REM started automatic repointing, and from 06:13:20 UT the burst location
stayed at the center of camera's field of view.
The observational conditions at that time of $\gamma$--trigger were suboptimal.
The burst occurred at a zenith distance $\approx68^{\circ}$, the sky was bright
due to a nearly full moon, and a large part of camera field of view had been
covered by the REM dome.  The Fig. 1 shows a summary of TORTORA,
 "Pi of the Sky" and Swift BAT light curves.
\begin{figure}
  \includegraphics[height=.3\textheight]{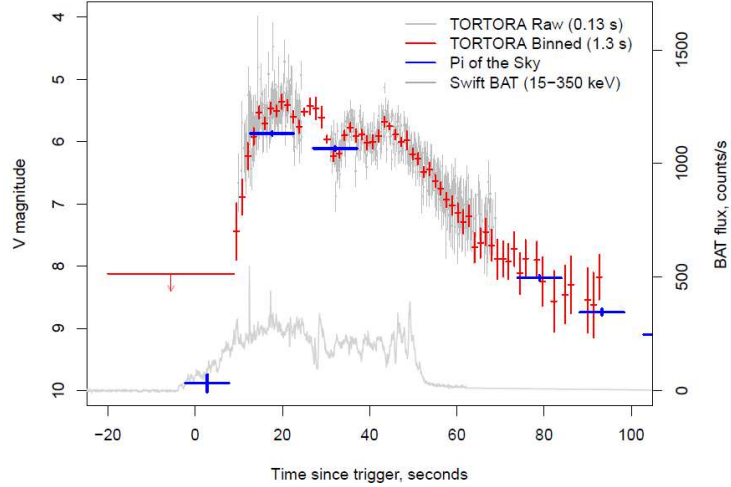}
  \caption{The light curve of GRB080319B acquired by TORTORA's wide-field
    camera (upper curve) alongside with Swift$/$BAT $\gamma$-ray one
    (lower curve, light gray color).  Also, transient brightness measurements by "Pi
    of the Sky" optical camera are shown.  The Swift$/$BAT light curve is the
    sum of all four energy channels. TORTORA data points show both full
    resolution (original data frames, grey color) and low-resolution (10 images
    co-added, red color). Full-resolution data are unavailable for the time interval of REM
    telescope repointing due to massive blurring of object image.}
\end{figure}

\subsubsection{Data Reduction}
The values of the effective air mass at middle time exposure and the seeing as
measured by La Silla--Meteo Monitor\footnote{www.ls.eso.org/lasilla/dimm/.} were
$\sim$ 2.6 and $\sim$ 0.9", respectively.  According to \citep{Sch98} Galactic
Extinction value is E (B--V) = 0.011 mag which (assuming $R_{v}$ = 3.1) implies
$A_{B}$ = 0.049, $A_{V}$ = 0.034, $A_{R}$ = 0.030.  The TORTORA fast wide-field
camera took the data on the field of GRB 080319B from 05:46:22 UT ($t_{0}=26$
minutes before trigger) to 06:15:41.00 ($t_{0}=172$ sec after trigger)
collecting $\sim$ 13320 unfiltered images with an effective exposure time of
0.13 sec without any temporal gaps between  consecutive frames.  From 23
sec up to 30 sec after the burst, the REM-telescope was repointing to the
location of GRB 080319B after the alert message was received through the GCN
notices by the Swift-BAT instrument.  The raw images stored in RAID have been
processed at the day time by a pipeline including TV-CCD noise subtraction,
flat-fielding to compensate vignetting due to objective design, and custom
aperture photometry code taking into account non self-averaging pixel
statistics caused by the image intensifier tube. Circular aperture photometry
was performed with \texttt{PHOT/DAOPHOT} function in
\texttt{IRAF}\footnote{IRAF: Image Reduction and Analysis Facility,
  http://iraf.noao.edu/.}.
For the REM repointing time interval fluxes have been derived using custom
elliptic aperture photometry code after summation of 10 consecutive frames with
compensated motion of the stars. Unfortunately, it seems impossible to
reconstruct the light curve of this interval with any better resolution due to
massive blurring of star PSF caused by their motion.  For all other intervals,
photometry has been performed both with 10-frames (1.3 sec effective exposure)
binning, and with original (0.13 sec) time resolution.
 As a result, we have identified a new peak in the prompt optical light curve.  Finally the photometry performed in
instrumental system was calibrated towards the V magnitudes of several nearby
Tycho2 stars.  We have no data on prompt light curve color for GRB 080319B at early
times ($t_{0}$ < $t$ < 60 sec).  Thus, no additional color corrections
have been applied to TORTORA's unfiltered data.

\section{Light Curve Structure}
TORTORA has tracked a fast rise of optical emission from $t_{0}$ +10 sec to
$t_{0}$ +15 sec, followed by a complex evolution until $t_{0}$ +45 sec and a
slow decay thereafter.  The rise from V $\approx$ $7.5^{m}$ to V $\approx$
$5.5^{m}$ may be approximated by a $\sim$ $t^{4}$ power-law originated at
$t_{0}$ $\approx$ 0; while $\gamma$--ray emission started earlier, at $t_{0}$
$\approx$ -4 sec.  The decay since $t_{0}$ +45 sec is a $\sim$ $t^{-4.6}$
power-law.  Four peaks can clearly be seen in optical data with an inter-peak
separation of $\sim$ 8.5 sec.

\subsection{Observed and Intrinsic Optical Parameters}
The fluxes of the four well-detected optical peaks $F_{opt,\; 1,\; 2,\; 3,\; 4}$
\begin{center}
\begin{equation}
F_{opt}= 827 \times (3.60 \times 10^{-9} \times 10^{-0.4 \times mag})
\end{equation}
\end{center}
were obtained by using the calibration of \citep{Fuk95} and were corrected for
galactic extinction (Table 1).  Host galaxy reddening correction $A_{v}$ was applied
assuming the mean value reported by \citep{Rac08}, E (B--V) = 0.05.
\begin{table}
\centering
\begin{tabular}{lrr}
\hline
   \tablehead{1}{r}{b}{Peak Flux }
  & \tablehead{1}{r}{b}{[\emph{erg  cm$^{-2}$ s$^{-1}$}]$\;\;\;\;\;$}
  \\
\hline
$F_{opt,\;1}$&($2.20$ $\pm$   $0.16)\times 10^{-8}$\\
$F_{opt,\;2}$&($2.18$ $\pm$   $0.28)\times 10^{-8}$\\
$F_{opt,\;3}$&($1.49$ $\pm$	$0.10)\times 10^{-8}$\\
$F_{opt,\;4}$&($1.61$ $\pm$   $0.10)\times 10^{-8}$\\
\hline
\end{tabular}
\caption{Prompt Optical Parameters of GRB 080319B: Peak Flux.}
\label{tab:a}
\end{table}
The isotropic equivalent Luminosity $L_{opt}$ (Table 2) for the well-detected
 optical peaks is related to the  peak  flux $F_{opt,\; 1,\; 2,\; 3,\; 4}$
using the equation
\begin{center}
\begin{equation}
L_{opt}=4 \pi \kappa_{opt}(z) D_l^2(z)F_{opt}
\end{equation}
\end{center}
where $D_l^2(z)$ is the luminosity distance for the cosmological standard model
and $\kappa_{opt}(z)$ is the cosmological $\kappa$-correction that accounts for
the transformation of the $V$ passband in the proper GRB frame:
\begin{center}
\begin{equation}
\kappa_{opt}=\frac{\int_\frac{\nu_{V_{0}}}{(1+z)}^\frac{\nu_{V1}}{(1+z)} \nu^{-\beta} d \nu}{\int_{\nu _{V_{0}}}^{\nu_{V1}} \nu^{-\beta} d \nu}= \frac{1}{(1+z)^{1- \beta}}
\end{equation}
\end{center}
Here, $\nu_{V_{0}}$ and $\nu_{V_{1}}$ are the frequency boundaries of the $V$
band and $\beta$ is the power-law index in the optical spectrum
$F_{\nu} \propto \nu^{- \beta}$.
For $\kappa$-correction we assume $\beta = 0.50 \pm 0.07$  as reported by \cite{Rac08}.
\begin{table}
\centering
\begin{tabular}{lrr}
\hline
   \tablehead{1}{r}{b}{Peak Luminosity}
&  \tablehead{1}{r}{b}{[\emph{erg s$^{-1}$}]$\;\;\;\;\;\;\;\;\;\;\;\;$}
  \\
\hline
$\;\;\;\;\;\;\;\;\;\;$$L_{peak,\;1}$&$(7.24$ $\pm$ $0.71)\times 10^{49}$\\
$\;\;\;\;\;\;\;\;\;\;$$L_{peak,\;2}$&$(7.21$ $\pm$ $1.08)\times 10^{49}$ \\
$\;\;\;\;\;\;\;\;\;\;$$L_{peak,\;3}$&$(4.91$ $\pm$ $0.42)\times 10^{49}$\\
$\;\;\;\;\;\;\;\;\;\;$$L_{peak,\;4}$&$(5.30$ $\pm$ $0.45)\times 10^{49}$\\
\hline
\end{tabular}
\caption{Prompt Optical Parameters of GRB 080319B: Peak Luminosity.}
\label{tab:a}
\end{table}
The optical fluence $S_{opt,\;V}$ was determined by numerically integrating the
prompt light curve in the interval from the earliest observation to the
latest one  with a power-law interpolation of the flux in the segments between
the experimental points yielding $S_{opt,\;V}$ = $(7.17$ $\pm$
$1.80)\times10^{-7}$ $erg$ $cm^{-2}$ ($1.87$ sec < $t$ < $86$ sec).
The isotropic equivalent total energy in $V$ band $E_{opt,\;V}$ in the rest
frame of the source was determined from the optical fluence $S_{opt,\;V}$
using the relation:
\begin{center}
\begin{equation}
E_{opt,\;V} = \frac{4 \pi \kappa_{opt}(z) D^2_l(z) S_{opt,\;V}}{(1+z)}
\end{equation}
\end{center}
yielding $E_{opt,\;V}$ = $(1.21$ $\pm$ $0.30)\times 10^{51}$ $erg$.
The complex evolution from $t_{0}$+15 sec to
$t_{0}$+45 sec consists of two  well-distinct regions of different  peak intensity levels:
$F_{opt,\;1,\;2}$ $\approx$ $2.19$ $\times$ $10^{-8}$  $erg$  $cm^{-2}$ $s^{-1}$
from $t_{0}$+15 sec to $t_{0}$+30 sec and
$F_{opt,\;3,\;4}$ $\approx$ $1.55$ $\times$ $10^{-8}$  $erg$  $cm^{-2}$ $s^{-1}$ from
$t_{0}$+30 sec to $t_{0}$+45 sec which roughly correspond to two regions of
$\gamma$-ray light curve.
The event is extreme not only in observed properties but also intrinsically: it
is the most luminous event ever recorded at optical wavelengths and has an
exceedingly high isotropic-equivalent energy release in $\gamma$-rays. The
previous record was held in brightness by GRB 990123 \citep{Ake99}, GRB 050904 \citep{Taglia}, GRB 061007 \citep{Mundell}.
In spite of its initial brightness, the behavior of the afterglow at
middle/late time does not appear to be peculiar.  The extrapolated luminosity
e.g. at 10 hour and at 13 hour after the trigger in the rest frame of the
source ($L_{opt@10^{h}}$ $\approx$ 3.16 $\times$ $10^{44}$ $erg$ $s^{-1}$ and $L_{opt@13^{h}}$
$\approx$ 9.77 $\times$ $10^{43}$ $erg$ $s^{-1}$, respectively) are comparable with the average
luminosity of the afterglow sample detected over the past few years
\citep{Nar08}.

\subsection{Variability Analysis}
As reported previously the light curve is approximated by a four nearly equidistant flares with an
inter-peak separation of $\sim$ 8.5 sec in observer frame.  The Power Density Spectrum between 10
sec and 50 sec confirms this feature with a 99.999\% confidence.  When we use
the high resolution data from $t_{0}$+40 sec to $t_{0}$+50 sec a periodicity at
a frequency of $\sim$ 0.9 Hz is detected with a 99\% confidence and
it may be interpreted as a Lense-Thirring precession or nutation \citep{Beskastro}.

\section{Conclusions}
The prompt optical emission of the GRB 080319B is peculiar for several reasons:
it is the most luminous event ever recorded reaching a  visual peak absolute
magnitude of $M_{V,\;peak}$ = -38.4 and is the variable object at cosmological
distances with the shortest optical periodicity ever discovered.  The temporal
structures reflect the behaviour of the internal engine (periodicity of $\sim$
8.5 seconds for overall emission, four peaks) and the disk precession ($\sim$ 1.1
second period on the last peak), and imply the newborn stellar-mass black hole
accreting from massive disk as an internal engine of this burst.

\begin{theacknowledgments}
  This work was supported by the University of Bologna Progetti Pluriennali
  2003, by grants of CRDF (No. RP1-2394-MO-02), RFBR (No. 04-02-17555 and
  06-02-08313), INTAS (04-78-7366), and by the Presidium of the Russian Academy
  of Sciences Program.

\end{theacknowledgments}



\bibliographystyle{aipproc}   

\bibliography{proc_B}

\hyphenation{Post-Script Sprin-ger}
\begin{thebibliography}{19}
\expandafter\ifx\csname natexlab\endcsname\relax\def\natexlab#1{#1}\fi
\providecommand{\enquote}[1]{``#1''}
\expandafter\ifx\csname url\endcsname\relax
  \def\url#1{\texttt{#1}}\fi
\expandafter\ifx\csname urlprefix\endcsname\relax\def\urlprefix{URL }\fi
\providecommand{\eprint}[2][]{\url{#2}}

\bibitem[Akerlof et~al.(1999)]{Ake99}
C.~Akerlof, R.~Balsano, S.~Barthelmy, J.~Bloch, P.~Butterworth, D.~Casperson,
  T.~Cline, S.~Fletcher, F.~Frontera, G.~Gisler, J.~Heise, J.~Hills, R.~Kehoe,
  B.~Lee, S.~Marshall, T.~McKay, R.~Miller, L.~Piro, W.~Priedhorsky,
  J.~Szymanski, and J.~Wren, \emph{Nature} \textbf{398}, 400--402 (1999).

\bibitem[Gehrels et~al.(2004)]{Geh2004}
N.~Gehrels, G.~Chincarini, P.~Giommi, K.~O. Mason, and J.~A. Nousek, \emph{et
  al., The Astrophysical Journal} \textbf{611}, 1005--1020 (2004).

\bibitem[Piccioni et~al.(1993)]{picc}
A.~Piccioni, A.~Bartolini, C.~Cosentino, A.~Guarnieri, S.~R. Rosellini,
  A.~di~Cianno, A.~di~Paolantonio, C.~Giuliani, E.~Micolucci, and
  G.~Pizzichini, \enquote{An updating about FIP: a photometer devoted to the
  search for optical flashes from gamma-ray bursters,} AIP Conference
  Proceedings 280, American Institute of Physics, New York1st Compton Gamma Ray
  Observatory Symp, 1993, pp. 1152 -- 1155.

\bibitem[Beskin et~al.(1999)]{Besk99}
G.~M. Beskin, V.~Plokhotnichenko, C.~Bartolini, A.~Guarnieri, N.~Masetti,
  A.~Piccioni, A.~Shearer, A.~Golden, and G.~Auriemma, \emph{Astronomy and
  Astrophysics Supplement} \textbf{138}, 589--590 (1999).

\bibitem[Karpov et~al.(2005)]{KarpFAV}
S.~Karpov, G.~Beskin, A.~Biryukov, S.~Bondar, K.~Hurley, E.~Ivanov, E.~Katkova,
  A.~Pozanenko, and I.~Zolotukhin, \emph{Il Nuovo Cimento C} \textbf{28},
  747--751 (2005).

\bibitem[Molinari et~al.(2006)]{Mol06}
E.~Molinari, S.~Bondar, S.~Karpov, G.~Beskin, A.~Biryukov, E.~Ivanov,
  C.~Bartolini, G.~Greco, A.~Guarnieri, A.~Piccioni, F.~Terra, D.~Nanni,
  G.~Chincarini, F.~M. Zerbi, S.~Covino, V.~Testa, G.~Tosti, F.~Vitali, L.~A.
  Antonelli, P.~Conconi, G.~Malaspina, L.~Nicastro, and E.~Palazzi, \emph{Il
  Nuovo Cimento B} \textbf{121}, 1525--1526 (2006).

\bibitem[Racusin et~al.(2008{\natexlab{a}})]{Rac08_GCN}
J.~L. Racusin, N.~Gehrels, S.~T. Holland, J.~A. Kennea, C.~B. Markwardt,
  C.~Pagani, D.~M. Palmer, and M.~Stamatikos, \emph{GRB Coordinates Network,
  Circular Service} \textbf{7427}, 1 (2008{\natexlab{a}}).

\bibitem[Golenetskii et~al.(2008)]{Gol08}
S.~Golenetskii, R.~Aptekar, E.~Mazets, V.~Pal'shin, D.~Frederiks, and T.~Cline,
  \emph{GRB Coordinates Network, Circular Service} \textbf{7482}, 1 (2008).

\bibitem[Vreeswijk et~al.(2008)]{Vree08}
P.~M. Vreeswijk, B.~Milvang-Jensen, A.~Smette, D.~Malesani, J.~P.~U. Fynbo,
  P.~Jakobsson, A.~O. Jaunsen, and C.~Ledoux, \emph{GRB Coordinates Network,
  Circular Service} \textbf{7451}, 1 (2008).

\bibitem[Wo\'{z}niak et~al.(2009)]{Woz08}
P.~R. Wo\'{z}niak, W.~T. Vestrand, A.~D. Panaitescu, J.~A. Wren, H.~R. Davis,
  and R.~R. White, \emph{The Astrophysical Journal} \textbf{691}, 495--502
  (2009).

\bibitem[Cwiok et~al.(2008)]{Cwi08}
M.~Cwiok, W.~Dominik, G.~Kasprowicz, A.~Majcher, A.~Majczyna, K.~Malek,
  L.~Mankiewicz, M.~Molak, K.~Nawrocki, L.~W. Piotrowski, D.~Rybka,
  M.~Sokolowski, J.~Uzycki, G.~Wrochna, and A.~F. Zarnecki, \emph{GRB
  Coordinates Network, Circular Service} \textbf{7439}, 1 (2008).

\bibitem[Karpov et~al.(2008)]{Kar08}
S.~Karpov, G.~Beskin, S.~Bondar, C.~Bartolini, G.~Greco, A.~Guarnieri,
  D.~Nanni, A.~Piccioni, F.~Terra, E.~Molinari, G.~Chincarini, F.~M. Zerbi,
  S.~Covino, V.~Testa, G.~Tosti, F.~Vitali, L.~A. Antonelli, P.~Conconi,
  G.~Cutispoto, G.~Malaspina, L.~Nicastro, E.~Palazzi, E.~Meurs, and
  P.~Goldoni, \emph{GRB Coordinates Network, Circular Service} \textbf{7452}, 1
  (2008).

\bibitem[Schlegel et~al.(1998)]{Sch98}
D.~J. Schlegel, D.~P. Finkbeiner, and M.~Davis, \emph{Astrophysical Journal}
  \textbf{500}, 525--553 (1998).

\bibitem[Fukugita et~al.(1995)]{Fuk95}
M.~Fukugita, K.~Shimasaku, and T.~Ichikawa, \emph{Publications of the
  Astronomical Society of the Pacific} \textbf{107}, 945--958 (1995).

\bibitem[Racusin et~al.(2008{\natexlab{b}})]{Rac08}
J.~L. Racusin, S.~V. Karpov, M.~Sokolowski, J.~Granot, X.~F. Wu, and
  V.~Pal'Shin, \emph{et al., Nature} \textbf{455}, 183--188
  (2008{\natexlab{b}}).

\bibitem[Tagliaferri et~al.(2005)]{Taglia}
G.~Tagliaferri, L.~A. Antonelli, G.~Chincarini, A.~Fernández-Soto, D.~Malesani,
  M.~Della~Valle, P.~D'Avanzo, A.~Grazian, and V.~Testa, \emph{et al.,
  Astronomy and Astrophysics} \textbf{443}, L1--L5 (2005).

\bibitem[Mundell et~al.(2007)]{Mundell}
C.~G. Mundell, A.~Melandri, C.~Guidorzi, S.~Kobayashi, I.~A. Steele,
  D.~Malesani, L.~Amati, P.~D'Avanzo, D.~F. Bersier, A.~Gomboc, and E.~Rol,
  \emph{et al., The Astrophysical Journal} \textbf{660}, 489--495 (2007).

\bibitem[Nardini et~al.(2008)]{Nar08}
M.~Nardini, G.~Ghisellini, and G.~Ghirlanda, \emph{Monthly Notices of the Royal
  Astronomical Society} \textbf{386}, 87--91 (2008).

\bibitem[Beskin et~al.(2009)]{Beskastro}
G.~Beskin, S.~Karpov, S.~Bondar, A.~Guarnieri, C.~Bartolini, G.~Greco, and
  A.~Piccioni, \emph{eprint arXiv:0905.4431, submitted}  (2009).

\end{thebibliography}

\IfFileExists{\jobname.bbl}{}
 {\typeout{}
  \typeout{******************************************}
  \typeout{** Please run "bibtex \jobname" to optain}
  \typeout{** the bibliography and then re-run LaTeX}
  \typeout{** twice to fix the references!}
  \typeout{******************************************}
  \typeout{}
 }

\end{document}